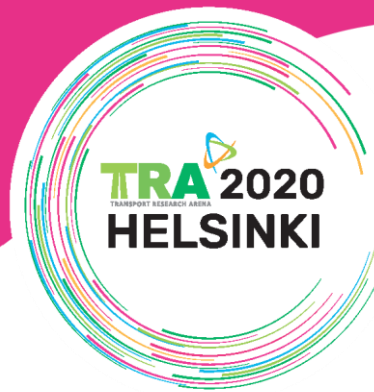

# Gamification and engagement of tourists and residents in public transportation exploiting location-based technologies

**Bruno Cardoso[a]\*, Miguel Ribeiro[a], Catia Prandi[b], Nuno Nunes[a]**

[a]*ITI Larsys, Caminho da Penteada, Funchal, 9020-105, Portugal // Tecnico Lisboa, Av. Rovisco Pais 1, Lisboa, 1049-001, Portugal*
[b]*Università di Bologna, Via Zamboni, 33, 40126, Italy*

Abstract

Cities are becoming very congested. There is a need to reduce the number of private cars on the roads, by maximising the potential for local public transport.
With the increasing awareness of transport that is sustainable in the sense of environmental impact, but also climate and social, there is the need to create engagement into public transportation. Gamification, which is the use of game elements in non-game contexts, has proven to deliver very positive results, by turning regular activities into engaging ones, which are fun to perform. We have designed a mobile application, that interacts with short-range wireless communication technologies, inviting people to use public transport. To evaluate the solution, we have created a questionnaire based on the System Usability Scale, but also using usability testing with specific tasks.

\* Corresponding author. Tel.: +351-291-721-006;
  *E-mail address:* bruno.cardoso@tecnico.ulisboa.pt

# 1. Introduction

With cities becoming more and more crowded, there is a need to reduce the number of private cars driving on the road. For this, the solution is to maximise the potential for local public transport, by making it attractive, and a convenient alternative, but also by integrating the public transport system with people's lives, creating an appealing means of transportation, improving the urban environment. People are still not adhering to public transport. Only 9.1% do Europe is using the bus on regular usage.

According to the sustainable mobility paradigm by D.Banister(2008), should we understand Transport as a derived demand or as a valued activity? The new standard regarding conventional transport analysis is based on the premise that travel is a cost, and that travel times should be as short as possible. However, this is changing as new technology allows much greater travel time flexibility, including mobile working. What this is confirming is that we can introduce new mechanisms to induce behavioural change in mobility, by creating engagement to this means of transportation, but also by developing secondary activities, creating more sustainable environments.

With computing, technology became cheaper and more user-friendly. Connected computing technology enables new services and interaction between users and the service provider, such as a new type of service called Mobility-as-a-Service (MaaS), as explained by P. Jittrapirom-et-al. (2017). MaaS can be thought of as a new transport solution. According to S.Hietanen(2014), MaaS is a mobility distribution model that deliver users' transport needs through a single interface of a service provider. This is one of the reasons behind Uber, that offers a platform that matches travellers demanding a trip and car owners that want to supply this trip, by A.Hagiu(2014). Multi-Sided Platforms (MSPs) reduce search and transaction costs, as in S.Hientanen(2014).

It is common to have the transport network connected via GSM devices, sharing information that can be worked and delivered to the final user, or even to evaluate the current state of the system. When combined with mobile devices, these advanced are most valuable for the users, as they can schedule their lives knowing precisely at what time should they expect their transport. This adds value to the product, by keeping users informed at no cost, but also by making them adhere to the system.
Public transport companies are already using this regularly, yet, although keeping the user informed, this fails to create engagement with the user and the transport, especially regarding customer loyalty.

# 2. Gamification

The use of newly developed words to reference new concepts might bring some confusion, especially when the term is related to a well-established word. This is one of the problems with the idea of Gamification. The close relation to gaming tends to create a mixture where the user believes that game, gaming and gamification are the same things. The idea might have similarities, but it is essential to accurately establish what each concept means, and how not to mix them.

*2.1. Gamification*

The concept of Gamification is something that is finally starting to become recurrent. The term is relatively new, in Epstein(2013), but the idea of resembling games mechanics and techniques to engage with users and to solve a problem is already being used at least since 1960, in G.Boulet(2016), A.Marczewski(2017), S.Nicholson(2015) and Zichermann and Cunningham(2011). Although not addressed as a concept, the idea was discussed in several publications since then, such as in Gartner(2011), and even portrayed in different Hollywood releases. It was not until 2003 that the term Gamification started having an impact as a whole. Nick Pelling is known for coming up with this term. About seven years later, and while gaining interest, its definition can be defined as "The process of game-thinking and game mechanics to engage users and solve problems" in Zichermann and Cunningham(2011). This definition might seem vague, but the basic idea behind it is to make it as uncertain as possible, but always related to game design techniques. The idea is to bring all the different solutions developed in games for non-gaming contexts. Being this such a broad range, it is expected to bring several different views into gamification, while research in behavioural psychology confirms the success of the various solutions, just like the success of social games, as explained by Huotari and Hamari(2012). As a result, by making it a game, and rewarding the user at the end, we are facing positive behaviour change. By escalating this to society in general, we amplify its growth, as V.Kalantzis(2017).

It is also the term that we use when a couple is preparing to spend the rest of their life together. It might seem out of the blue, but the fact is that engagement is the period at which a person has a great deal of connection with another person, place, idea or thing. Although there is no specific metric to evaluate engagement, we can use a couple of different parameters to conclude engagement. For example, using Recency, Frequency, Duration, Virality and Ratings, we can derive an engagement score for this. According to each business type, there might be score fluctuations, as some business types focus more on Frequency and Recency (for example a Restaurant) but less about the Duration. The old idea of pushing customers to buy more is slowing disappearing, as the primary goal is to generate more revenue, and not worry so much about sales as a whole. This is why engagement does not follow revenue; instead, behind engagement, revenue follows, Zichermann and Cunningham(2011).

Gamification allows for greater motivation when executing a specific task. Motivation is also described as being a technical approach to define "behaviour" which represents the reasons for people's actions, desires and needs, as in Ziesener-et-al. (2013).

*2.2. Game Elements and Dynamics*

To use game mechanisms, it is required to elaborate on game elements and dynamics, as described by Sailer(2017). These are the layers that define a game, the pieces that all together result in a game. The art of game design, by J.Schell(2014) classifies four central themes that define a game. These are Mechanics, a Story, the Aesthetics and the Technology. It is essential to combine all four of them, without prioritising any of them, as they are all critical to the final result, J.Schell(2014), McGonigal(2011).

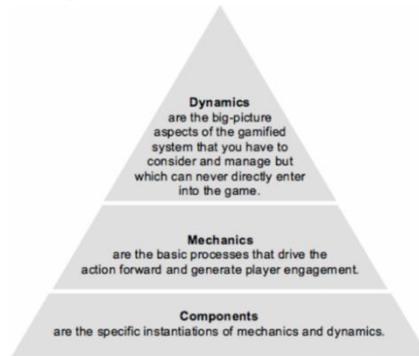

Fig 1: The Game mechanisms pyramid

The Aesthetics are what the user sees, how the game looks, sounds, and feels. This is the most crucial aspect of game design since the aesthetics have a direct relationship to the player's experience. It works with the mechanics of the game, as the mechanics of the game should be able to support the visual part, by amplifying and reinforcing it. As described by S.Deterding et al. (2011):

"You will want to choose mechanics that make players feel like they are in the world that the aesthetics have defined, and you will want a story with a set of events that let your aesthetics emerge at the right pace and have the most impact " The aesthetics are also responsible for creating and reinforcing a memorable experience. Technology Not the generic term technology, we refer to technology here as the materials and interactions that make the game possible, from the traditional pen, pencil board games, to the high-powered lasers and hardware. Technology is what blends the player with the game, allowing an entirely new set of game-play mechanics.

Components make up the largest group of game elements. In many ways, the components are the more specific form of either Dynamics or Mechanics. These elements are less abstract than the first two categories and lead to actual tools that can be employed to begin to incorporate gamification into the environment of interest.

The implementation of gamification follows a progression of steps, in K.Werbach(2012):

- 1. Define business objectives
- 2. Delineate target behaviours
- 3. Describe players
- 4. Devise activity cycles
- 5. Incorporate
- 6. Deploy the appropriate Mechanics and Components

## 3. Technologies

There are different approaches available to mobile development, and different technologies involved, including some that are more focused on gaming, while others are broader. Choosing the right technology is crucial to the success of the project.

*3.1. Web applications*

There are different available options for mobile development. When developing an application with a specific objective, it is required to analyse the different options available further, so that the result is versatile, fast and supportable, by being upgradable.

There are three primary development methodologies in regards to mobile applications, all of them with their advantages, and this is why it is so important to analyse the application's requirements, to find the methodology that best suits its purpose. The choice will vary with the target audience, budget and range of devices the app will support. Mobile WebApp, as in Charland(2011), also known as Progressive Web Apps are applications that look and feel like native mobile applications, that typically run on any browser. The advantage is that they are mostly generic, having almost no compatibility issues due to their nature. As a downside, they tend to lack hardware compatibility and are limited to the browser's capacities. In regards to native applications, they have the most significant advantage in any project as these are the most efficient regarding speed and hardware usage. This kind of development requires the official SDK for each platform, and each application has its programming language. As such, it is time savoury when developing to more than one Operative System. Then, there is a new methodology called hybrid applications. These are developed using a framework, which generally uses standard web technologies like HTML, CSS and Javascript to pack and deploy an application into a mobile device. These are then uploaded to the different Application Stores like Apple Store or Play Store.

There are several available options to develop such a project. While most of them have progressed well, being highly customizable and using external resources, native development is of the least desirable choices, for the increased difficulty when developing to 2 different platforms, the case with this project. Since cross-platform development tools are now jointly running at the same pace as native development app, there is an increased advantage in using them. For the different options available, the fact that React Native is composed of declarative components, and uses the same building blocks that both Android and iOS users, made us choose this technology for the application.

## 4. Implementation

This project takes place in Madeira Island, Portugal. More specifically, in the city of Funchal, one of the six European touristic sites where Civitas Destinations exist. Madeira is known for tourism since the XVII century, and has a particular attention to sustainable mobility, as in Prandi et al. (2017). It first started as a therapeutic destination for the British people, not only for its landscapes but also for its weather, in A.Vieira(2019).

In the first proposal, there was the intention of integrating short-range wireless communication technologies, based on radio signals. These would be used as locators for the application by creating interaction with the user. At the time, three different technologies were discussed. The project would either beneficiate from Near Field Communication(NFC), the communication protocol that works at a short distance, Radio Frequency Identification(RFID), which resembles NFC, except that it does not have to be in a line of sight, and the iBeacon protocol. This technology would have to be compatible with react native, reliable and also work on different environments such as high-temperature scenarios, crowded places or high precision location situations. Although the first two were interesting, they would require interaction between the users mobile phone and the NFC/RFID tag, making it unreliable in case the user was far away from the tag. The decision was for the iBeacon protocol, as in N.Newman(2014), but a modified version, with added features. Developed by Estimote, the estimote proximity and sticker beacons, are relying on location and feature exclusive cloud access to make bulk changes to every beacon. Ranging from 7 to 30 meters, the beacons also include NFC, temperature and motion sensors. The usage of external hardware is crucial to connect the application(user) to the location(beacon). Just as planned, we are using Bluetooth Low Energy Beacons to assist on adding interaction in the app. As soon as this was intended, there were several distance tests planned, including indoors and outdoors.

*4.1. BLE device capability in the bus environment*

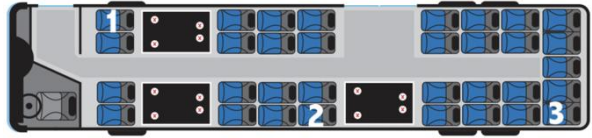

Fig 2: Different scanned locations inside the bus

These tests happened in 2 different days, with two different bus types and manufacturers. This was only possible due to the cooperation of the local bus company. The beacons were installed at the bus station so that nobody would tamper with each beacon inside the bus. For each beacon location, there were two beacons, one sticker and one beacon. The test would compare both beacon emissions at a particular time, being both at the same specific place. Using a beacon scanner, the vast amount of data would then be analysed to understand how the beacons reacted to the environment.

The results were positive. For the estimote proximity beacons, there is a significant number of broadcast beams for the observation time. For a 24-minute journey, with different bus occupancies, there is an average of 200 broadcasts received from the beacons. This is independent of the place of observation inside the bus. With each trip having a duration of between 20 and 40 minutes, having between 7.50 and 10 broadcasts per minute is more than sufficient to detect that the player is inside the bus, and heading to the place where the adventure takes place. As for the low power estimote stickers, there is a limited broadcast number, and are very dependent on the player's position inside the bus. At the farthest place from the beacon, there is an average of 0.16 broadcasts per minute, and this may vary with the number of passengers inside the bus.

It is worth mentioning that the environment influences a lot on the propagation conditions. Even replicating the same scenario, with the same people inside the bus, not only the beacons vary on strength, but also external factors influence. This is the reason for the mixed results for some locations.

As these results show, if there is passive tracking inside the bus, then it is required to have at least two estimote stickers inside the bus, or one estimote proximity beacon since these have a stronger signal.

*4.2. MARGe - the name*

Of several different names and acronyms that were thought to this project, the one that obtains consensus was MARGe. The idea was to have a name that connected with the island of Madeira, but also that reflected gamification. MadeiRA Gamified experience was the result. MARGe has a meaning and a display name for the application.

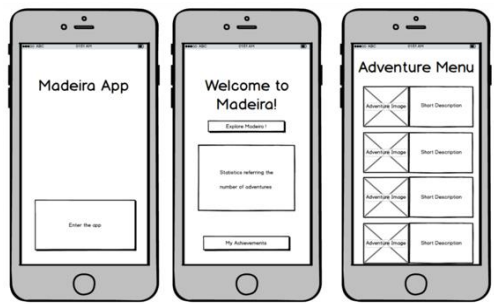

Fig 3: Low Fidelity prototype

From the time the first mockup of the application appeared, there were many changes, not only visually, but also in terms of backend and functionality. The early versions of the mockups were developed to understand if all the requirements and features made sense on a low fidelity prototype. Through development, some initial ideas had to be changed, since there were limitations to the platform in terms of external hardware access and graphical development. These low fidelity prototypes were very useful to understand the proper way to integrate external hardware. Since the early beginning, different characteristics from these prototypes were maintained and ported to the target product. At the time, these prototypes were developed using the wireframing tool Balsamiq.

From this early mockup, the app was ported to non-functioning high fidelity prototype using Adobe XD. The focus on these first mockups would be to add the early implementations of gamification, focusing on obtaining awards and transforming the standard user path into an adventure, resembling a scavenger hunt.

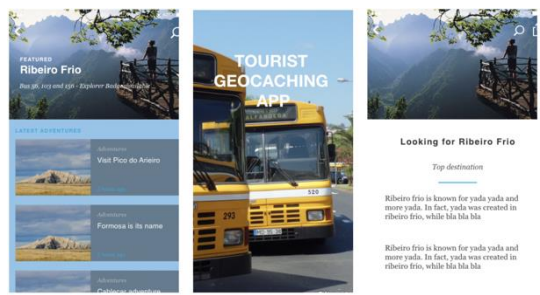

Fig 4: First Adobe XD mockups

*4.3. Features*

After several months of work, and different iterations over different backends and components, the first release candidate of MARGe is ready. MARGe is an android and iOS application, compatible with Android 8 and iOS 11 and above. MARGe integrates a secure authentication process, based on Googles Firebase Authentication. The app is consistent with the Estimote proximity beacons and also with the iBeacon protocol. The app is programmed to load all content from the internet, to make sure it is easily updatable, without requiring a new installation. This is also an advantage of the size of the application. MARGe is also based on gamification techniques, and, as such, motivates users to continue using the application, by creating engagement. Most design features in MARGe follow the material design guidelines. The application was conceived to have two different Approaches. A Benefits Approach, which focuses on communicating value or specific benefits to new users for the first time they open the application, and an Interactive Approach, which allows users to participate in the onboarding process by employing an interactive app onboarding flows, in Csikszentmihalyi(2008). By doing this, we always make sure the users learn about the app while interacting with it.

*4.4. Game Elements*

The developed application uses game elements from the three different levels of the pyramid previously described. There are two dynamics in the application. They influence the gamification experience, despite not being directly visible. These are:
- Progression (players evolution) - by providing an awards page, summarising how each award was obtained and giving hints on how to obtain new ones.
- Narrative(a form of teaching lessons) - by making each adventure a story, made of different stages, that further explain the adventures surroundings.

The mechanics part is responsible for creating engagement with the users. These are also used to achieve elements of dynamics. These are:
- Rewards (obtained for performing tasks) - By proving badges for each adventure, but also with specific badges that reinforce the usage of the application.
- Resource acquisition (obtaining items) - By creating different easter eggs in the application, that only get triggered if the user explores the application.
- Challenges (puzzles or other tasks that require effort to solve) - By implementing a set of clues to obtain the answer and advance onto the next stage, but also by implementing a quiz in each adventure, that requires effort to complete successfully.

Finally, components are the visibility of the elements. These are the state the elements take. These are:
- Levels (defined steps in players progression)
- Badges (visual representation of achievements)
- Quests (predefined challenges with objectives and rewards)
- Points (earned by completing tasks)
- Achievements (objectives completed)
- Leaderboards (visual representation of players progression in comparison to others)

Using different elements, such as the leaderboard, can accomplish and encourage competition among most users. A top leaderboard can promote the satisfaction of competence, but also autonomy for most users.

Another vital engagement factor is the score. Giving points for each finished adventure, not only relates with the leaderboard, but also start creating engagement with the user, by triggering a factor of addiction to the application, and also related to a sense of achievement.

Lastly, we also use badges to represent achievements, as these are delivered when a user completes an adventure, but also when obtaining 100% at a quiz or finding an easter egg. The application begins with a splash screen showing the MARGe logo. While the logo is showing, the application is loading all required components for the application to run. After loading, moving to the next screen is as easy as pressing a button. This button will bring the user to the language selector, where there are four different choosable languages. Before loading the different languages, there is a loading screen overlay, where the app detects if the user is returning to the application, or if it is the first time for that user. If the user is running the app for the first time, the first screen after the splash screen is the language selector one. For the app to work reliably, we chose to use external authentication services, to have a secure, reliable service. This way, not only we get higher speeds when transitioning between the app and the authentication services, but we also reduce the app's payload. By simplifying these steps for the user, we are making the application more user-friendly, and accessible to a broader audience, including the elderly.

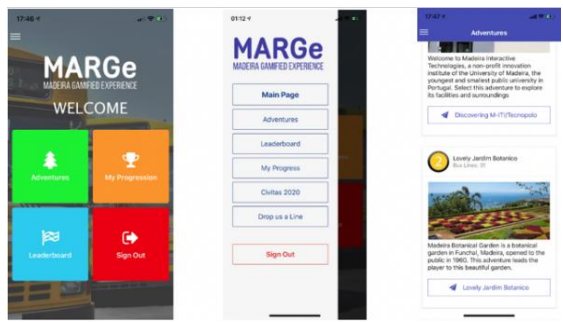

Fig 5: MARGe Application, main screen, side menu and adventure selector

If login is detected, then there is a database entry with the userId, and with that, the previously selected language. This is also useful when a user returns to the application since the language is automatically selected, without even showing the different options, bypassing the Welcome Instructions, Login/Register/Forgot Password screen, and on to the main page directly.

In the main page, the user is greeted in the chosen language and is also presented with four squared quick access menu buttons. The idea behind this menu is to access the main activities available in the application quickly. The secondary ones are available on the side menu. Of all the options, there is a quick access button to open the adventure selector, access the Users progression screen, a button for the leaderboard and one for the feedback screen. Also, from the side menu, there is a shortcut to the CiViTaS Destinations screen, showing institutional information. All of these screens are available on the main page, meaning that the user can navigate between all of them in both directions.

The adventure selector screen fetches a set of adventures from firebase, already in the pre-selected language. These are then displayed using card images, featuring a small text previewing the app, the award for that adventure and an image of the final place of the adventure. For each adventure, there is an identifier, an availability flag (if not available, the adventure selector menu alerts the user), the award id that adventure unlocks, the different bus lines, image, name and short description.

Choosing an adventure does not become a commitment for that adventure. The next screen explains every detail about the adventure, how many kilometres will the adventure cover, assuming the user starts at a common start point, any specific requirements for the adventure, and some pictures.

At the end of this screen, then, the user must confirm if it will be taking the adventure, and then, the application becomes one way only, until the end of the adventure. It is worth noting that the adventure will record a users stage in the adventure so that if the user exits the application when returning to that specific adventure, a prompt will ask if the user wants to resume. The same things happen when the user has already completed the adventure, encouraging users to try other adventures.

When in an adventure, there are four standard screens. These are interchangeable, and its content is adapted to the current state of an adventure. The first screen here is an Information screen, where a preview of the adventure is presented, coupled with some images and the final awards one may win at the end of the adventure. These are a different place- holders that are retrieved from our database, in a pre-selected language.

The second screen also has some information about the adventure, but, it also adds beacon location integration. The grey overlay showing an activity indicator transforms into a screen width button, unlocking access to the next screen as soon as the user is within range. This is part of the steps of the adventure later described in the database diagram.

The third screen is an example of the questionnaire screen. This is a multiple choice quiz, where answering right will add points to the overall score, but answering wrong will also subtract them. The minimum available score is zero, anything below that turns into zero so that it is not demotivational. This interactive quiz shows what would have been the right answer when the user fails. In the end, there is a screen showing the users score.

The fourth screen here is an example of a numbered steps screen. The idea here is to use it as a guide to accessing the last location. With that in mind, as soon as the right side column is scrolled, each number on the left column scrolls, and fills the circle, confirming to which text the number corresponds to.

As previously explained, this is just an example sequence, as each of these screens can be interchanged when programming the application.

### *4.5. Back-Office and storage scheme*

For the backend, We are using the firebase real-time database, in Li(2018). This NoSQL database is cloud-hosted, where data is stored as JSON and synchronised real-time to every connected client. This eases up synchronisation between different platforms. On Signup, each user received its unique User ID, to be able to store and retrieve data belong- ing to a specific user.

## 5. Evaluation

In this section, we evaluate whether the demonstrated development supports the proposed solution. We start by describing our evaluation process, where we combine different tests, a questionnaire and also a System Usability Scale, followed by an analysis of the results obtained.

For this, there are two different methods for testing: In the usability testing, we will be testing how hard it is for users to perform different use cases in the app, while the non-functional requirements testing will focus on the number of resources used by the app in different devices. We are also developing a System Usability Scale questionnaire to be completed after each practical test session.

### *5.1. Usability testing*

We will be using this section to detail every test and the procedures taken. We are also taking into account different factors for these tests, such as our target audience, to make sure the results are valid, and not showing a trend. For this, and since MARGe has a target audience of adults and young adults, we have defined our target audience as to be- ing composed of people of both genders aged between 17 and 55. Besides age, it is also important to understand everyone's background, ranging from people with a vast level of technological experience, but also those that are not that fond of technology.

### *5.2. Results and Discussion*

We have obtained eight different usability tests. For the different tasks, there is a big difference in the observed values and variation.

- For the first task, where the goal is to perform a simple action that requires pressing a button, the values range from 7 seconds to 32 seconds, where the average is of 15 seconds and the deviation of 8 seconds. There was only one error in 1 test, and it was unintentional. The user told the right name but mistook the button.
- For the second task, where the goal was to complete an adventure, going through every stage of the adventure, and waiting for the beacon detection, the results were positive. The average time for this task was of 4 minutes and 54 seconds, with a deviation of 1 minute and 33 seconds. For this test, there was only one error at one screen, when a user, after completing a question on the quiz, wanted to go back and change the selected option. This is a bug that should not happen, and the feedback was of high value.
- The third task was a simple one, yet, it leads to some error making. The task consisted of checking the overall percentage and score. This involved accessing the Awards screen. For most users, the task was completed successfully in no time, but for some users, this was a difficult task, leading to up to 2 errors. Some users opened every screen until finding the percentage, while others chose the right button in the first try. The average time for this task was of 16 seconds, with a deviation of 9 seconds. The average number of errors for this task was considered high, as it reached 0.8, with a deviation of 0.92. This was due to the users that searched for the feature, pressing the different buttons on the main page.
- For the last task, where the users had to look for the feedback screen, the average time for this task was of 1 minute and 38 seconds, with a deviation of 54 seconds. The feedback feature is regarded as a secondary feature, although important, and was moved to the side menu, instead of being on the main page. This was on purpose not to clutter the main menu. Due to this, the expected number of errors was higher than the previous tasks. The average number of errors for this task was of 1.7, with a deviation of 1.83. Most users completed the task successfully with zero errors, yet, users are making 2, 3 and 4 errors while looking for the feature.

We have also obtained eight different answers on the SUS, in Brooke(1996) and Lewis(2009). The System Usability Scale test has determined a percentile ranking with the value of 83, which, according to the general requirements for the SUS, is graded an A mark, being on the Promoter level, and also on the Acceptable level. As for an overall appreciation, these results are on the border between "Good" and "Excellent".

## 6. Conclusions

The biggest reason for this research was to use Gamification as a way to induce behavioural change in Mobility, by designing and testing mobility-related behavioural change applications, but also by promoting sustainable mobility, by making use of public buses, cycling or even walking. Of the many different gamification methods, our goal was to select the ones that would fit best on public transportation, but also by creating engagement with our users. The means to do this would also include providing an assisted guide to visit different places, using sustainable mobility, but with a new motivation to do so. By gathering award for the different places, the user visited, and also by progressing on the number of completed adventures and fostering competition for the leaderboard table. To accomplish this, we have developed a mobile application resourcing to external Bluetooth hardware, both compatible with the two major mobile Operating Systems, Android and iOS. This was made possible with React Native, by creating a custom solution for our specific situation. At the end of the development, we have tested and evaluated our study, by creating a process containing different usability testing, from 4 scripted usability tests, where we measure how long it takes the user to perform a specific task, the number of errors and any feedback they may give at a specific situation. Moreover, we also finish these tests with a System Usability Scale questionnaire to derive a usability score based on well-established usability criteria.

## 7. Declarations

*7.1. Availability of data and material*

The datasets and materials used and/analysed during the current study are available from the corresponding author on reasonable request.

*7.2. Competing interests*

The authors declare that they have no competing interests

*7.3. Funding*

This work was done under the CiViTAS Destinations project measure MAD6.1.

*7.4. Authors' contributions*

The first and second authors have made substantial contributions to the project, and work conception and all authors have drafted the work and revised it.

*7.5. Acknowledgements*

The authors would like to thank everyone involved in this project, ARDITI, ITI Larsys, Horarios do Funchal.